\documentclass[12pt,preprint]{aastex}
\usepackage{float}
\usepackage{arydshln}
\usepackage{xcolor}
\usepackage{graphicx}
\usepackage{natbib}

\newcommand{\Msun}{\hbox{M$_{\odot}$}}

\newcommand{\Mjup}{\hbox{M$_{\rm Jup}$}}

\newcommand{\ha}{H$\alpha$}

\newcommand{\galex}{{\sl GALEX}}

\shorttitle{ACRONYM of the $\beta$ Pictoris Moving Group}

\shortauthors{Shkolnik et al.}

\begin{document}


\title{All-sky Co-moving Recovery Of Nearby Young Members. (ACRONYM) II: The $\beta$ Pictoris Moving Group
\altaffilmark{1}\\}


\author{Evgenya~L.~Shkolnik}
\affil{School for Earth and Space Exploration, Arizona State University, Tempe, AZ, 85281 USA}
\email{shkolnik@asu.edu}

\author{Katelyn N. Allers\altaffilmark{2}}
\affil{Department of Physics and Astronomy, Bucknell University, Lewisburg, PA 17837, USA}

\author{Adam L. Kraus}
\affil{Department of Astronomy, The University of Texas at Austin, Austin, TX 78712, USA}

\author{Michael C. Liu\altaffilmark{2}}
\affil{Institute for Astronomy, University of Hawaii at Manoa, 2680 Woodlawn Dr., Honolulu, HI 96822, USA}

\and

\author{Laura Flagg}
\affil{Physics \& Astronomy Department, Rice University, 6100 Main MS-550, Houston, TX 77005, USA\\Department of Physics and Astronomy, Northern Arizona University, Flagstaff, AZ, 86011 USA}

\altaffiltext{1}{Based on observations made with the IRTF, Keck and Magellan/Clay telescopes.
}

\altaffiltext{2}{Visiting Astronomer at the Infrared Telescope Facility, which is operated by the University of Hawaii under contract NNH14CK55B with the National Aeronautics and Space Administration.
}

\begin{abstract}

We confirm 66 low-mass stellar and brown dwarf systems (K7-M9) plus 19 visual or spectroscopic companions of the $\beta$ Pictoris Moving Group (BPMG). Of these, 41 are new discoveries, increasing the known low-mass members by 45\%. We also add four objects to the 14  known with masses predicted to be less than 0.07$\Msun$.
Our efficient photometric+kinematic selection process identified 104 low-mass candidates which we observed with ground-based spectroscopy.  We collected infrared observations of the latest spectral types ($>$M5) to search for low gravity objects.  
These and all $<$M5 candidates were observed with high-resolution optical spectrographs  to measure the radial velocities and youth indicators, such as lithium absorption and \ha\ emission, needed to confirm BPMG membership, achieving a 63\% confirmation rate. 
We also compiled the most complete census of the BPMG membership with which we tested the efficiency and false-membership assignments using our selection and confirmation criteria.
We assess a group age of 22$\pm$6 Myr using the new census, consistent with past estimates. With the now densely sampled lithium depletion boundary, we resolve the broadening of the boundary by either an age spread or astrophysical influences on lithium burning rates. We find that 69\% of the now known members with AFGKM primaries are M stars, nearing the expected value of 75\%.
However, the new IMF for the BPMG shows a deficit of 0.2-0.3\Msun\ stars by a factor of $\sim$2. We expect that the AFGK census of the BPMG is \textit{also} incomplete, probably due to biases of searches towards the nearest stars.

\end{abstract}



\section{Introduction}\label{intro}

The discoveries of young moving groups (YMGs) throughout the past two decades have provided samples of young, nearby stars (e.g.~\citealt{ kast97,webb99,torr00,zuck00,zuck04,shko12,malo13,malo14,krau14,alle16}) for studies of star formation, stellar evolution, and activity-rotation-age relations. They also serve as prime targets for direct imaging searches for circumstellar disks, close stellar and brown dwarf (BD) companions and young exoplanets.  These groups most likely formed in coeval, co-spatial, and co-moving molecular clouds and then
spatially dispersed over time, now linked  through their common space motion and indications of youth.  Two of the youngest such groups are the 8--10 Myr (\citealt{barr06,bell15}) TW Hya Association (TWA; \citealt{webb99,zuck01}) named for the isolated T Tauri star, and the 20--25 Myr (\citealt{bink14,malo14,mama14,bell15}) $\beta$ Pic moving group  (BPMG; \citealt{zuck01b,zuck04,torr06,torr08,schl10,schl12a,schl12b}), named for its debris-disk and exoplanet hosting hot star.

Surveys for young and active stars have revealed $\approx$10 additional YMGs with ages between 8 and 300 Myrs, filling a critical age gap between the very young star-forming regions and the old field population \citep{mama16}. 
The close proximity to these YMG members to Earth ($\lesssim$100 pc) is also beneficial for studies in need of high angular resolution, such as circumstellar disk and planet imaging, and sensitivity to low-mass stellar and substellar companions. At these young ages, disks and planets are also at their brightest and more easily detectable, e.g. the well-studied AU Mic debris disk \citep{liu04a,bocc15} and substellar companions \citep{delo12,bowl13,bowl15b}. 

In this paper, we present the \textit{All-sky Co-moving Recovery Of Nearby Young Members} (ACRONYM) of the BPMG, the second in our series of YMG member searches, after our first paper on the 40-Myr Tuc-Hor YMG \citep{krau14}.  Early literature values of the age of the BPMG were $\approx$12 Myr for over a decade (Table 1 of \citealt{mama14}). More recent work including lithium and isochrone analyses have converged to an age of 23 $\pm$ 3 Myr \citep{bink14,bell15}. By 23 Myr, the stars will likely have already formed their giant planets and are in the process of forming terrestrial planets (\citealt{liss87,liss96,keny06}).   Currently, there are two directly imaged planets known around BPMG members: a $\approx7$ \Mjup\ planet around its namesake A0 star, $\beta$ Pic (\citealt{lagr10}), and a 2.5 \Mjup\ planet around the F0 star, 51 Eridani (\citealt{maci15}). There is also one known free-floating planetary mass L dwarf PSO J318.5338-22.8603 \citep{liu13,alle16a}.  Smaller planets have not yet been found around such young stars primarily due to the detection limitations of direct imaging instruments and radial velocity (RV) searches. We focused our search on new low-mass members (K7-M9; M$_*\lesssim0.7\Msun$) around which most planets form and contrast ratios are more favorable for directly imaging planets and low-mass companions.

Prior to this publication, there were 146 systems confirmed as BPMG members consisting of 16 systems with A or F star primaries, 33 with G or K primaries, 80 M star (M0-M6) systems, and 14 brown dwarfs ($\geq$M7), yielding a 61\%  M-star fraction of the AFGKM stars. (See Appendix for complete census.\footnote{Three BPMG members do not have published spectral types.}) 
Yet, with M dwarfs making up 75\% of the stellar mass function in the field \citep{boch10}, we expected that $\sim$75 of the low-mass members of BPMG had yet to be discovered. In this paper, we report the confirmation of 66 low-mass BPMG members,  41 of which are new discoveries (37 M stars + 4 BDs), increasing the known M-star fraction to 69\%.


\section{Photometric Candidate Selection}\label{selection}

\begin{figure}[tbp]
\includegraphics[width=5.5in,angle=90]{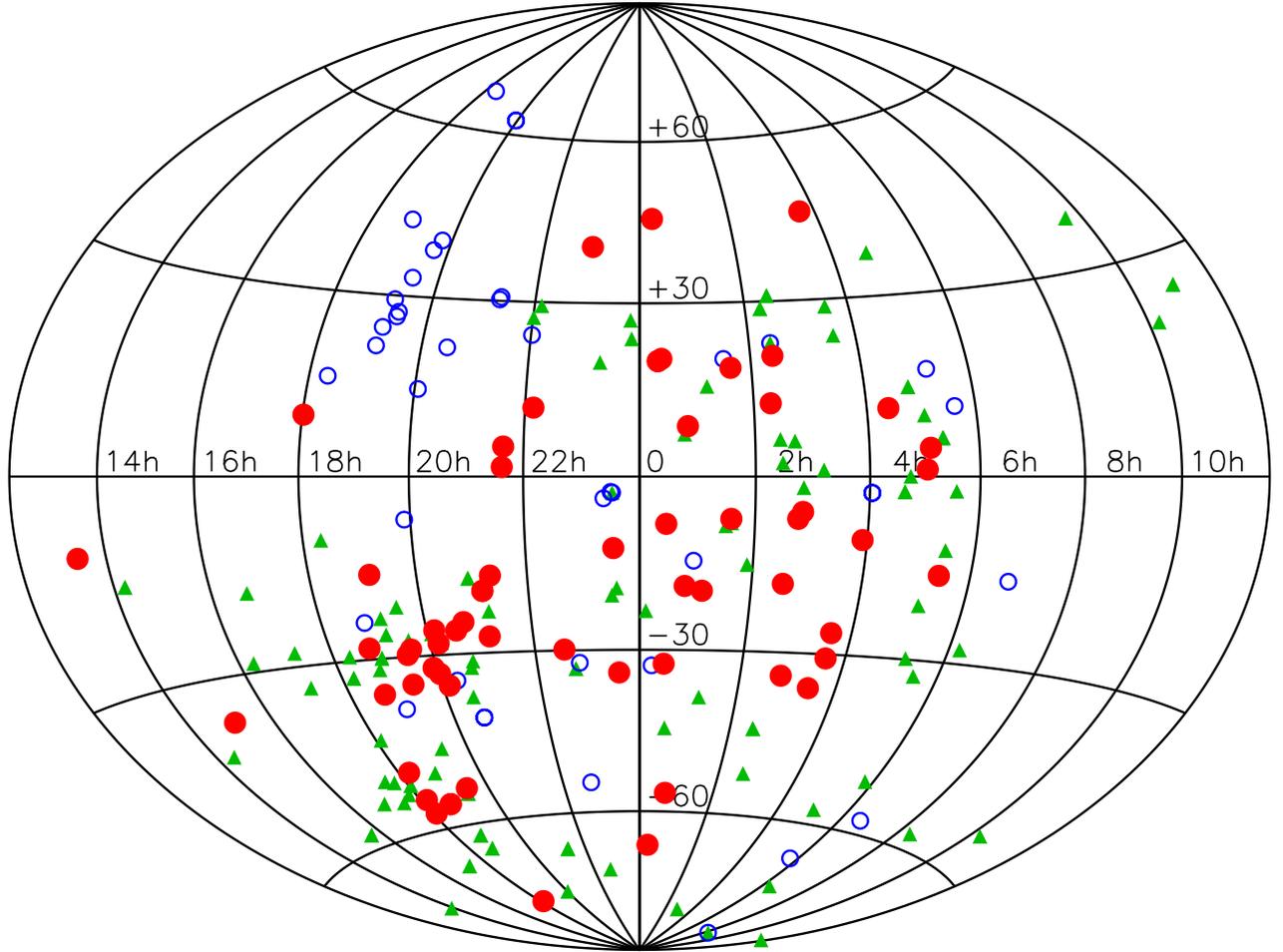}
	\caption{Positions of our candidate BPMG members on the sky shown in Aitoff projection. Candidates confirmed to be bona fide members (with both `Y' and `Y?' membership assessment in Table~3) are shown with red filled circles, while interloper field stars are shown with blue open circles. Known members from the literature (Appendix) are shown with filled green triangles. 
	\label{bpmgmap}}
\end{figure}

Our photometric selection process to identify BPMG candidates and the spectroscopic analysis needed for confirmation of new members are nearly identical to those used in our search for new Tuc-Hor moving group members \citep{krau14}. To  summarize, we combined astrometry and photometry from USNO-B1.0 (\citealt{mone03}), 2MASS (\citealt{skru06}), SDSS DR9 (\citealt{ahn12}), DENIS (\citealt{epch94}) and UCAC3 (\citealt{zach10}). These data yielded proper motions from the astrometry, and spectral types (SpT) and bolometric fluxes from SED fitting of the photometry for all sources in those catalogs. To concentrate our search on the most probable candidates, we narrowed our input sample using two spatial cuts. First, to avoid crowding and its effects on the inferred stellar properties, we neglected sources near the galactic plane and center ($|b| < 5^o$, or $|b| < 10^o$ and $310^o < l < 50^o$). Second, to focus our consideration to the locus of known BPMG members (Figure~\ref{bpmgmap}), we disregarded targets with $7h < \alpha < 18h$ . 

Nearly all of the sources that passed our two selection criteria have proper motions or color-magnitude diagram positions inconsistent with membership in the BPMG, so they were winnowed to find the small number of bona fide members. We first computed the kinematic distance modulus ($DM_{kin}$) that would minimize the difference between a source's observed proper motion and the expected proper motion for a BPMG member at that position on the sky. The magnitude of this discrepancy is reported as $\Delta_{PM}$. We identified 4486 sources where the observed and expected proper motions agreed within 3$\sigma$ (i.e., $\Delta_{PM} / \sigma_{\mu} < 3$) or the total discrepancy was $\Delta_{PM} < 10$ mas/yr, and  the inferred kinematic distance was $d \le 65$~pc ($DM_{kin} \le 4.0$). We then estimated the height above the main sequence by comparing the kinematic distance modulus to the spectrophotometric distance modulus, requiring $0.5 \le (DM_{kin} - DM_{phot}) \le 4.0$ to select pre-main sequence stars. The HR diagram criterion reduced our target list to 660
photometric+kinematic candidates. Finally, we narrowed our focus to
the 104 candidates with SpTs of K7--M9.
The SpT for each star was measured from the SED fitting process (SpT$_{SED}$) described by \cite{krau14}. In that paper, we showed that they are consistent with SpTs measured from optical spectra for K7 $\le$ SpT $\le$ M5.5.  Candidates with SpT$_{SED}$ later than M5 were targeted with IR spectroscopy (Section~\ref{ir_spectra}) to search for indications of low-gravity. Those candidates that appeared low-gravity and all those with SpT$_{SED}$ $<$ M5 were observed with optical spectroscopy. We measured the spectroscopic SpTs (SpT$_{spec}$) from the IR and/or optical data when possible, which are usually consistent to within one subclass for stars later than K7.  In a few cases, poor seeing did not allow for a reliable narrow-band TiO index to be measured (\citealt{shko09b}).  

Several studies have demonstrated that new members of YMGs also can be identified using ultraviolet photometry from \galex\ \citep[][]{find10,shko11a,rodr11,rodr13}.   A test of this selection process in our Tuc-Hor search \citep{krau14} showed that applying a \galex\ selection criterion to candidates is indeed more efficient, leading to a higher confirmation rate for spectroscopic follow-up. However, \galex\ only provides coverage of 2/3 of the sky (excluding the galactic plane), so $\approx$33\% of unknown members would not be confirmed.   Our photometric+kinematic selection procedure is unbiased toward stellar activity. Although the process requires the collection of more spectra of candidates,  it will ultimately discover 50\% more members with a small reduction in efficiency, i.e.~78\% with \galex\ pre-selection compared to $\approx$67\% without for the Tuc-Hor sample. 

We list the 104 late-type candidate members of the BPMG in Table~1 with their relevant photometric and kinematic information, as well as a log of the spectroscopic follow-up observations.

\section{Near-Infrared Spectroscopy of Late-M Candidates}\label{ir_spectra}

After photometric candidate selection, spectroscopic observations are necessary to assess youth. Spectroscopic youth indicators include features of low gravity, strong \ha\ emission, and Li absorption. For our faintest and latest SpT candidates, we first vetted for signs of low-gravity with lower-resolution near-IR spectra before collecting high-resolution optical spectra.

We obtained near-IR spectra of the 43 candidates with SED-fit spectral types $\gtrsim$ M4 using SpeX, the facility spectrograph of the NASA Infrared Telescope Facility (IRTF) on Mauna Kea.  We searched for signatures of low gravity in their spectra and re-measured their SpTs to confirm those determined by SED fitting using the method of \citet{alle13}, hereinafter AL13.  Our near-IR and SED spectral types agree well (to within one subclass). Spectra were reduced using Spextool\footnote{Version 3.4 for all but 2M00274534-0806046 which used Version 4.2} (\citealt{cush04}), which includes a correction for telluric absorption \citep{vacc03}.  

\begin{figure}[tbp]
\includegraphics[width=5.5in,angle=0]{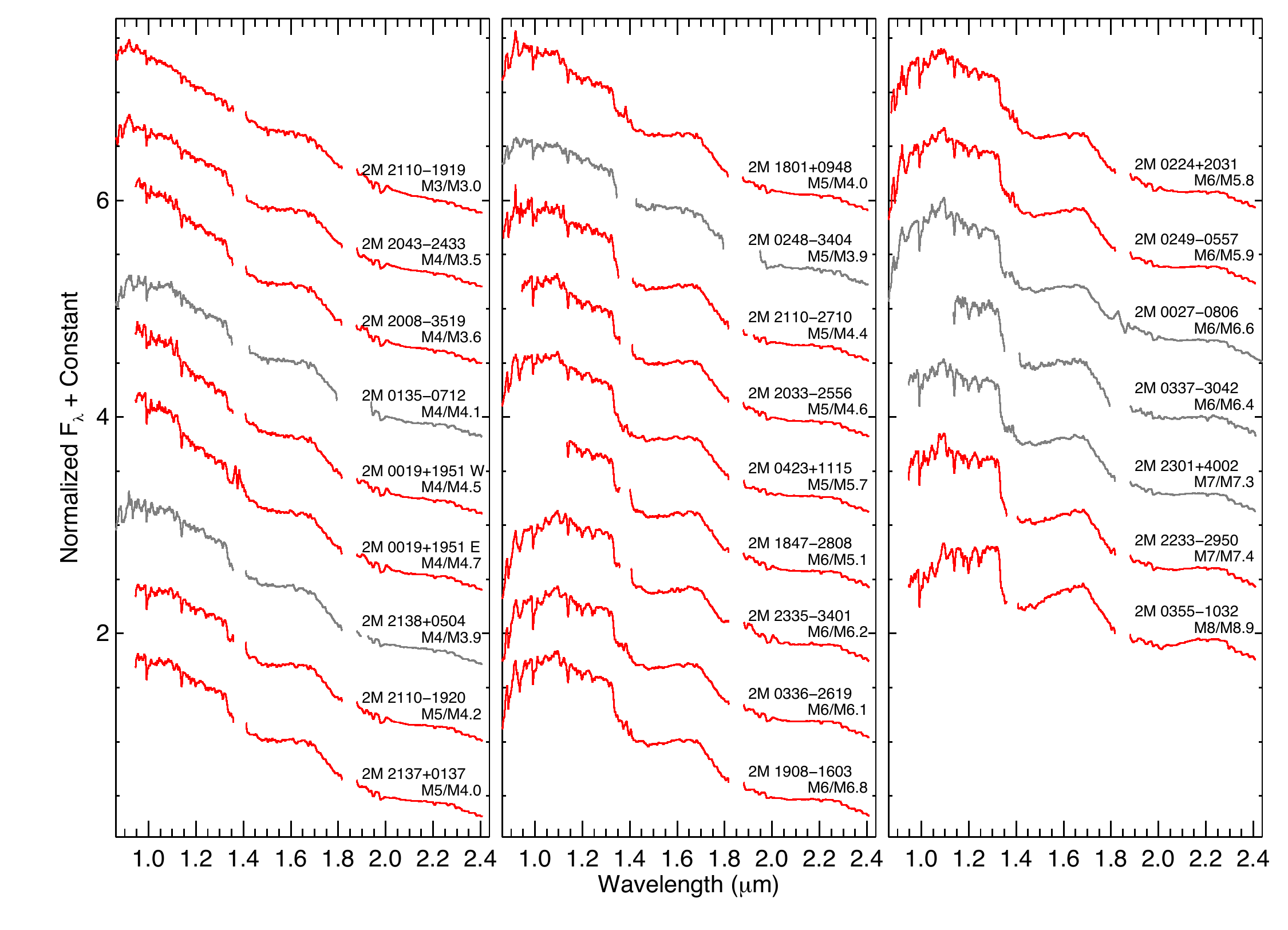}
	\caption{SpeX spectra of BPMG members.  All spectra have been smoothed to a resolution, $\lambda/\Delta \lambda~=~200$.  Objects that we confirm as members of BPMG are plotted in red.  Likely members (with a `Y?' designation in Table 3), are plotted in grey.  For each object the near-IR spectral type and SED spectral type are labeled (SpT$_{NIR}$/SpT$_{SED}$), and agree to within one subclass. 
	\label{bpmg_nir}}
\end{figure}

We classified our spectra following the method of AL13.  We determined gravity-insensitive near-IR spectral types using visual comparison with spectral standards from \cite{kirk10} as well as the spectral indices from AL13.  
We then calculated the AL13 gravity-sensitive indices for each spectrum and determined gravity classifications. The results are listed in Table~2.  Fifteen targets had near-IR spectral types earlier than M5, for which the AL13 gravity classification cannot be applied. Of the remaining 27 targets with SpT $\ge$ M5, 17 candidates (64\%) exhibited signs of youth, i.e.~intermediate (INT-G) or very low gravity (VL-G). These targets were then observed with high-resolution optical spectroscopy as described below to measure \ha, Li and radial velocities (RV). Other than three with ambiguous designations, all were confirmed as BPMG members.  Our SpeX near-IR spectra are displayed in Figure~\ref{bpmg_nir}.

\section{Membership Confirmation with Optical High-Resolution Spectra}\label{optical_spectra}

\begin{figure}[tbp]
\includegraphics[width=6.5in,angle=0]{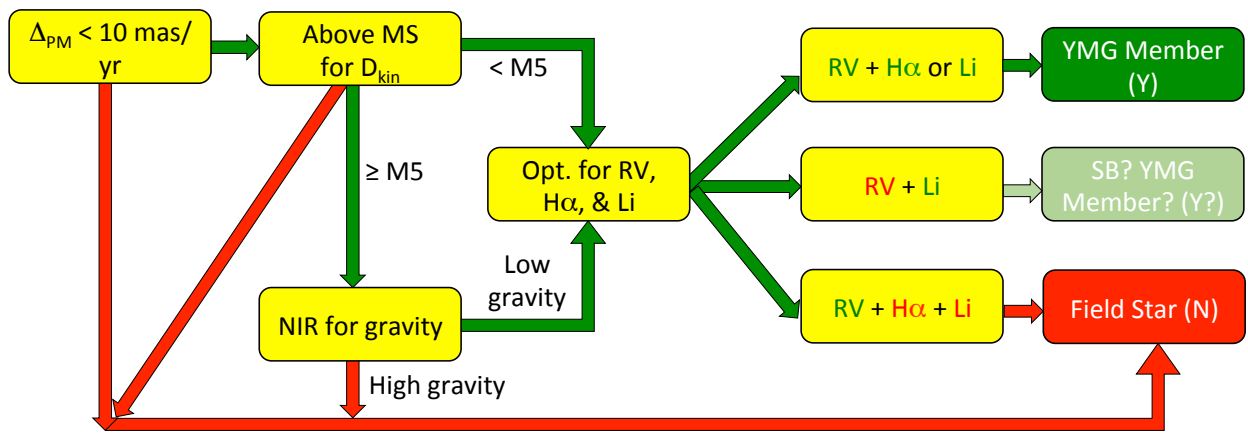}
	\caption{Flowchart for prescribing our young moving group selection process where red connectors and font denote ``No'' and green connectors and font denote ``Yes''.  An RV match is made when $|\Delta$RV$|$, the absolute difference between the kinematically predicted RV and the measured RV, is $<$5.4 km/s. An \ha\ match implies that the EW of the emission lies above the empirically determined limit derived by \cite{stau97b} for young ($\approx$50 Myr) clusters. The final status designations of `Y', `Y?' and `N' are listed for each candidate in Table~3. 
	\label{flowchart}}
\end{figure}

High-resolution spectra provide the RV necessary to calculate the space velocity (UVW), using the distance and proper motions, and thus to kinematically match the target to the moving group. With kinematic false membership rates of $\sim$50\% cumulatively for all YMGs and 6\% for the BPMG (e.g.~\citealt{pomp11,shko12}), both kinematics and independent spectral youth indicators are necessary to confirm a target as a bona fide YMG member. By combining all these data, we then designate each candidate in Table~3 with a membership status of `N' for a non-member when the kinematics or youth indicators conflict with membership, `Y' for confirmed members, and `Y?' for likely members, but whose radial velocities differ from the predicted values due to binarity in the system.  Our assessment procedure is summarized in Figure~\ref{flowchart}.
In this study, we used the kinematic distance derived by assuming a candidate is a group member and thus report the UVWs only for those which we assess to indeed be members. 

We acquired high-resolution optical spectra of our BPMG candidates, excluding those with IR spectra indicating field surface gravities, over  seven nights with Magellan Inamori Kyocera Echelle (MIKE) optical echelle spectrograph on the Clay telescope at Magellan Observatory and two nights with Keck's High Resolution Echelle Spectrometer (HIRES). (See Tables~1 and 3.)

The MIKE data were collected with the 0.5$\arcsec$
slit with a corresponding spectral resolution of $\approx$35,000 across the
4900--10000 \AA\ range of the red chip. 
The data were reduced
using the facility pipeline (\citealt{kels03}).\footnote{\url{http://code.obs.carnegiescience.edu/mike}} Each stellar exposure
was bias-subtracted and flat-fielded. After extraction of each order, the one-dimensional spectra
were wavelength calibrated with a ThAr arc lamp, which was taken after every stellar exposure to limit temperature differences and spectral drifts between the lamp and target exposures. To correct for any remaining 
instrumental drifts, we used the telluric O$_2$~A
band (7620--7660 \AA) to align the MIKE spectra to
$<$40 m/s. We then corrected for the heliocentric motion
of the Earth. The final spectra are of moderate S/N ($\approx$35 per pixel at 7000 \AA).

We used the 0.861$\arcsec$ slit with HIRES to give a spectral resolution of $\approx$58,000. The detector is a mosaic of three 2048 $\times$ 4096 15-$\micron$ pixel CCDs spanning 4900--9300 \AA\ and designated as the blue, green, and red chip. We used the GG475 filter with the red cross-disperser to maximize the throughput near the peak of an M dwarf spectral energy distribution. The HIRES data were reduced using the facility pipeline MAKEE\footnote{http://www.astro.caltech.edu/$\sim$tb/ipac\_staff/tab/makee/index.html} written by T.~Barlow. The MAKEE pipeline performs bias subtraction, flat-field correction, spectral extraction, wavelength calibration using the ThAr spectra, and a heliocentric velocity correction.  

On each night, we observed RV standards with spectral types spanning those of our BPMG candidates. The standards were taken from \cite{nide02}, with updated RVs if available from \cite{chub12}.   
Each night, spectra were also taken
of an A0V standard star for telluric line correction.  Spectroscopically derived SpTs (SpT$_{spec}$) are listed in Table 3.  For the late-type members we list the near-IR SpT. Optical SpTs were done with either TiO band indices or spectral line fitting following \cite{shko09b} and \cite{krau14}, respectively. In some cases due to poor continuum fitting, we include literature spectroscopic spectral types.

\subsection{Radial Velocities for Kinematic Matches to the BPMG}\label{rv}

Since our targets are relatively red, we limited our RV measurements to the reddest orders where the S/N is the highest. We cross-correlated each order between 7000 and 9000 \AA~(excluding those orders with strong telluric absorption) using the IRAF routine FXCOR
(\citealt{fitz93}). We measured the RVs from the Gaussian peak fitted to the cross-correlation function of each order and adopted the average RV of all orders with their standard deviation as the measurement uncertainty. The MIKE and HIRES data provide RV measurements to better than 1 km/s in almost all cases. 

\begin{figure}
\plotone{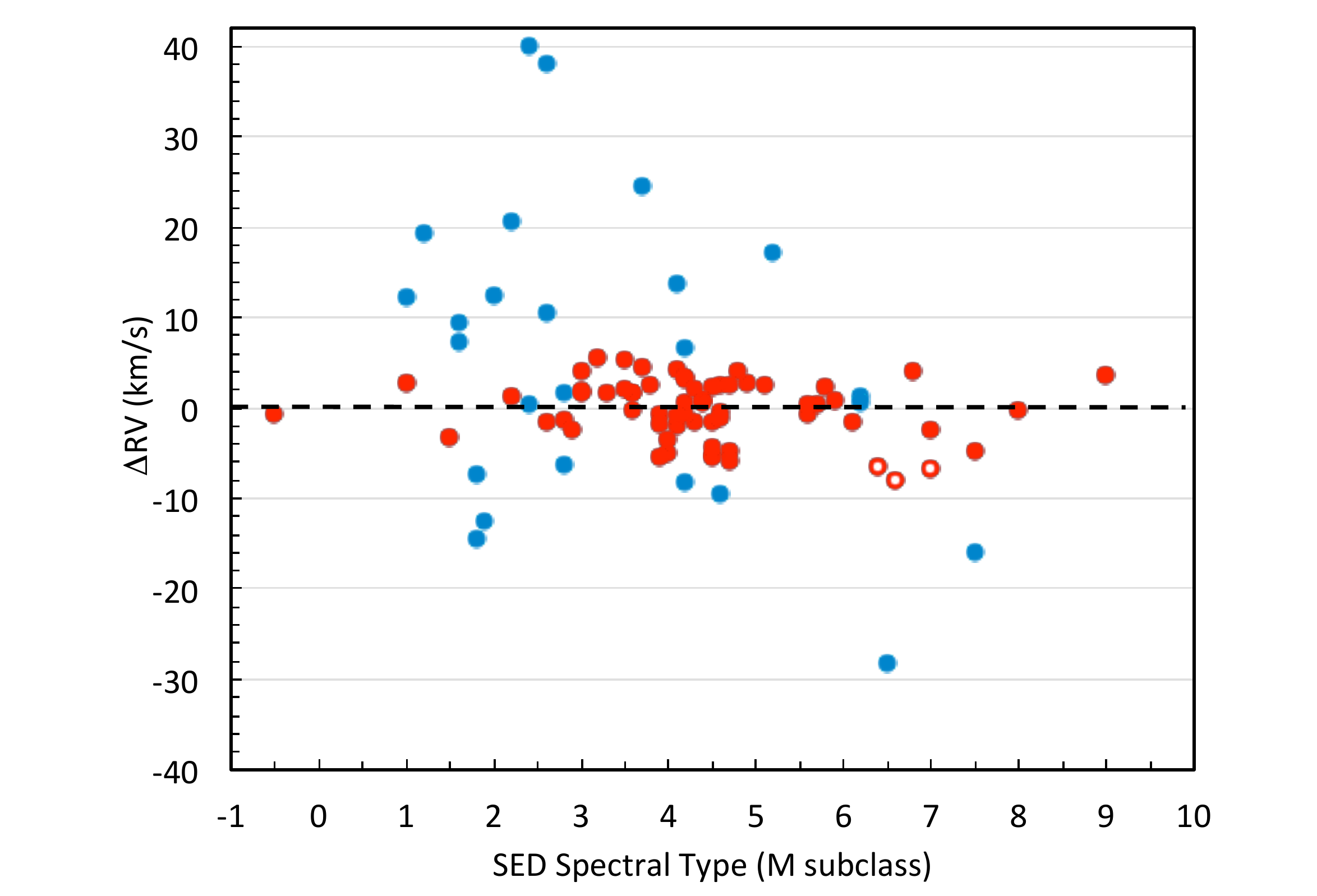}
\caption{RV difference ($\Delta$RV = RV$_{obs}$ -- RV$_{kin}$) for our targets observed with high resolution optical spectroscopy as a function of spectral type. We use the SpT$_{SED}$ for all targets with SpT$_{SED}$ $<$ M7. For objects $>$ M7, we plot the spectroscopically determined SpT listed in Table 3. As in Figure~\ref{bpmgmap}, members are denoted with filled red circles and nonmembers with blue circles.  Three possible members (with `Y?' designation shown as open red circles) have slightly discrepant RVs ($|\Delta$RV$| >$ 5.4 km/s), which may be due to unresolved binarity (SB1) but are confirmed to be young by the presence of Li absorption. These may also be young field interlopers. 
\label{spt_DRV}}
\end{figure}

The target RVs are listed in Table~3 along with the difference in measured and predicted RV, assuming the star is a BPMG kinematic match: $\Delta$RV~=~RV$_{obs}$~--~RV$_{kin}$. With an RV dispersion of known BPMG members of 1.8 km/s (\citealt{mama14}), 
we designate a star to be a kinematic match if $|\Delta$RV$| <$ 5.4 km/s, a 3$\sigma$ deviation from a perfect match (Figure~\ref{spt_DRV}). 

The high resolution of the optical spectra also allow for the identification of double- or triple-lined spectroscopic binaries (SBs). Ten of our targets are SBs, consisting of seven SB2s, two SB3s, and one SB1s (2MASS J02485260-3404246, SpT$_{SED}$=M3.9) which appears to be RV variable when compared to literature RV values.   Of the 10 SBs we identified (Table~3),\footnote{The fraction of 10\% SBs identified in this sample is lower than the 16\% found by \cite{shko10},  because the earlier study was pre-selected for high X-ray activity, biasing the sample towards tightly-orbiting, tidally spun-up SBs.} eight have systemic RVs consistent with BPMG members.

\subsection{\ha\ Emission and Lithium Absorption as Youth Indicators}\label{halpha}

Nearly all young stars with SpT $\ge$ M0 exhibit Balmer emission, most notably \ha. \cite{stau97b} demonstrated a clear demarcation between older M dwarfs and those younger than the $\approx$50 Myr old clusters, IC 2602 \citep{dobb10} and IC 2391 \citep{barr04}. Exceptionally strong \ha\ emission also can indicate that a young star still hosts a gas-rich protoplanetary disk, where accreting gas is falling along magnetic field lines from the circumstellar disk onto the star (e.g.~\citealt{barr03}). None of our targets showed signs of accretion, which is not unexpected at the BPMG's age.

\begin{figure}
\plottwo{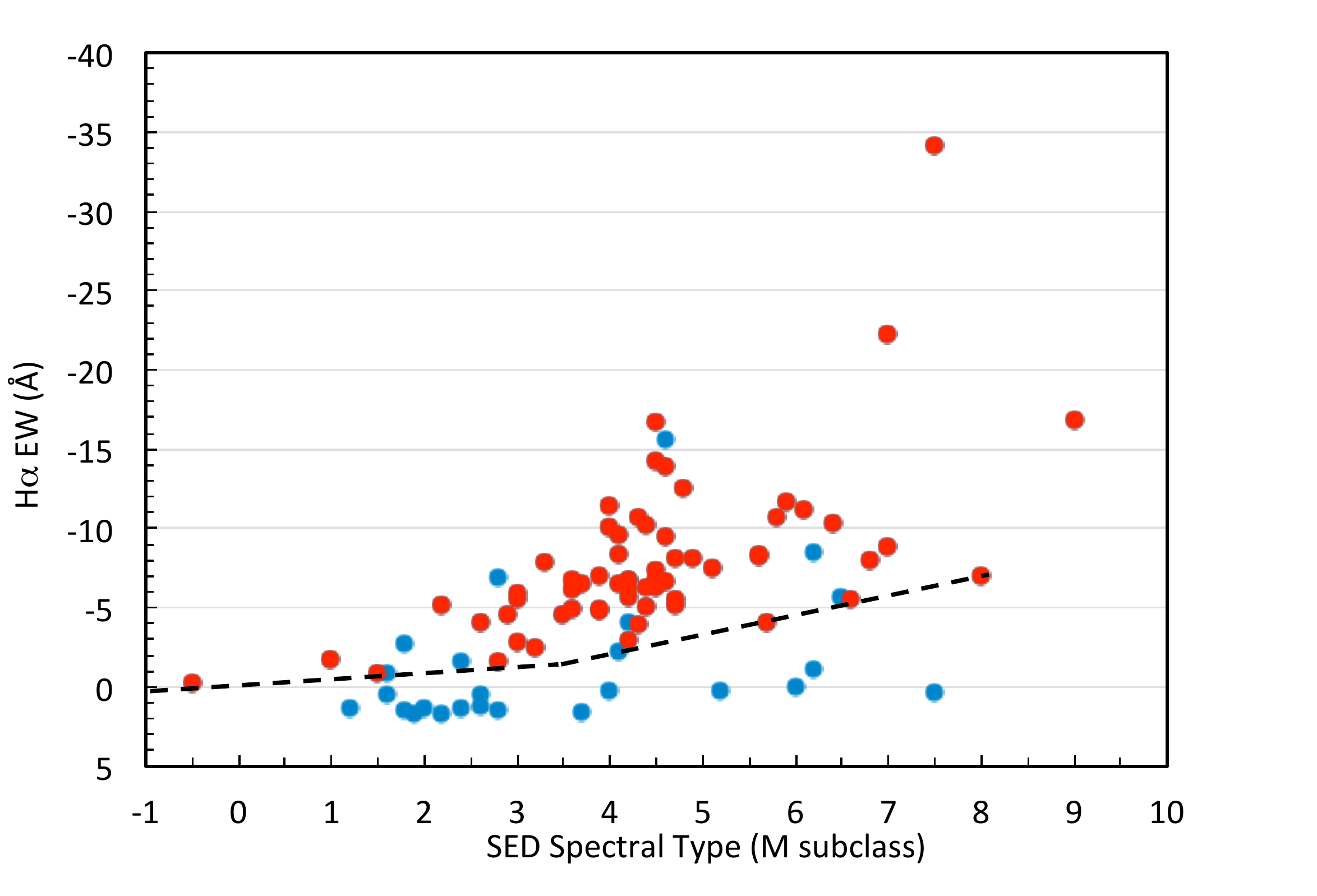}{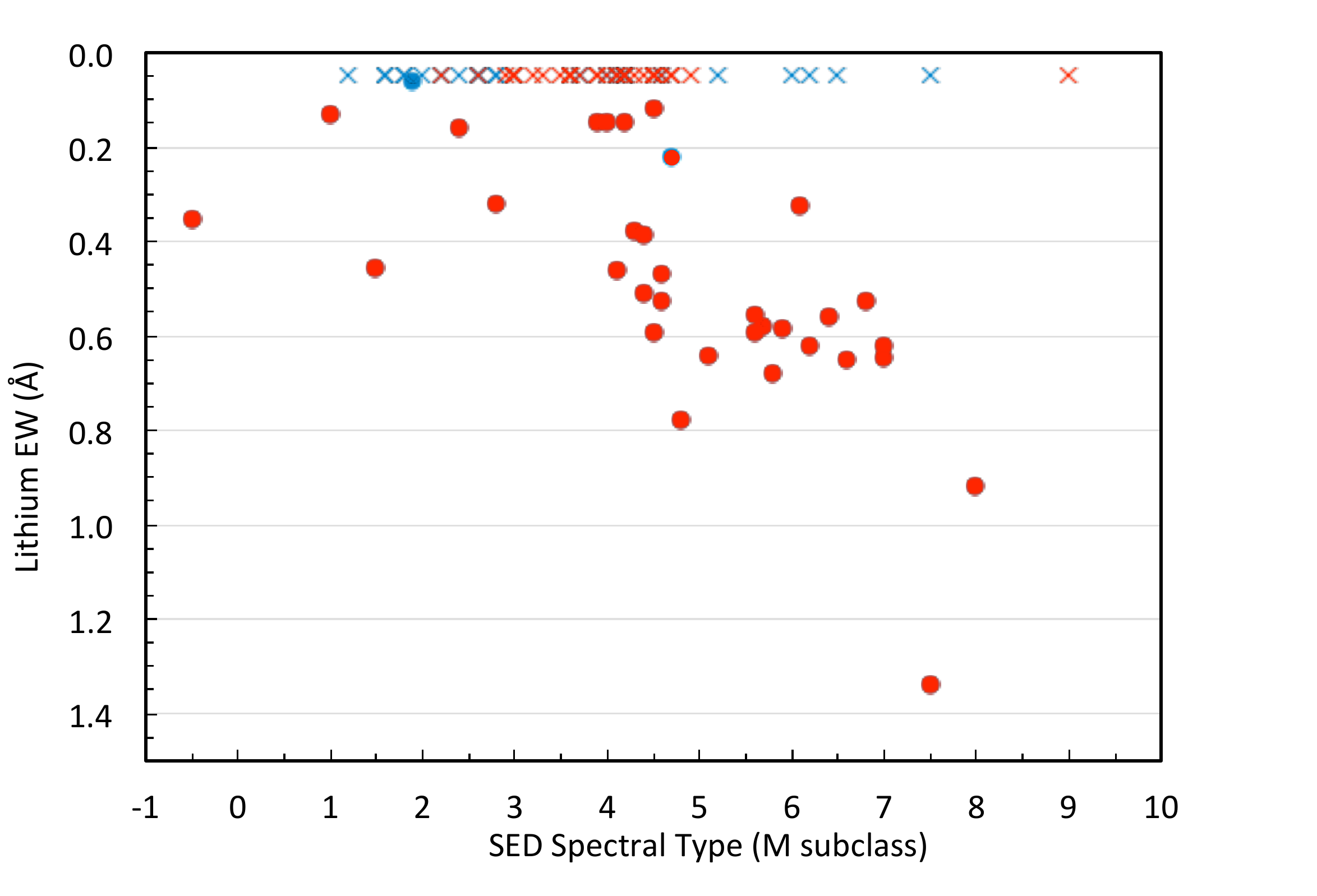}
\caption{Sample distribution of \ha\ emission (left) and lithium absorption (right) equivalent widths (EW). We use the SpT$_{SED}$ for all targets with SpT$_{SED}$ $<$ M7. For objects $>$ M7, we plot the spectroscopically determined SpT listed in Table 3.
The confirmed members are shown in red and the non-members in blue. The dashed line in the left plot represents the lower bound for \ha\ emission determined empirically by \cite{stau97b} from young ($\approx$50 Myr) clusters IC 2602 and IC 2391. Those cases with strong \ha\ emission but we do not identify as BPMG members have $|\Delta$RV$| >$ 5.4 km/s.  An `x' in the right plot represents an upper limit of 0.05 \AA\ for lithium EW. An analysis of the lithium depletion boundary and corresponding age for the BPMG is presented in Section~\ref{ldb}. \label{spt_halpha_lithium}}
\end{figure}

We apply the \cite{stau97b} \ha\ criterion to our sample to identify those with emission levels consistent with stars younger then $\approx$50 Myr. The \ha\ equivalent widths (EW) for our sample are plotted in Figure~\ref{spt_halpha_lithium} (left) identifying those stars we confirmed as BPMG members (as prescribed in Section~\ref{members} and Figure~\ref{flowchart}).  Those cases with strong \ha\ emission, but we do not identify as BPMG members, have $|\Delta$RV$| >$ 5.4 km/s.

Lithium (Li) absorption at 6708 \AA\ is another spectroscopic age indicator in the optical spectrum of pre-main sequence stars. Stars with M0 $<$ SpT $<$ M5 require 20--100 Myr to deplete their primordial lithium \citep{chab96}, with depletion time scales increasing for later SpTs. The earlier SpTs therefore allow for tighter age constraints. For objects with SpT $\ge$ M6, the detection of lithium sets an upper limit of $\approx$100 Myr on the star's age (\citealt{chab96,stau98}).  We use the detection of lithium to be a solid indication of youth.  However, the lack of lithium absorption in the spectra of early Ms is not confirmation of old age because the depletion timescales are short, providing an age limit comparable to the age of the BPMG. 
We measured lithium EWs or set limits\footnote{The lithium abundances have not been corrected for possible contamination with the Fe I line at 6707.44~\AA. Uncertainties in the setting of continuum levels prior to measurement induce EW errors of about 10-20~m\AA\/ with a dependence on the S/N in the region. We therefore consider our 2$\sigma$ detection limit to be 0.05 \AA.} for all of our targets for which we acquired optical spectra (Figure~\ref{spt_halpha_lithium}, right). An analysis of the lithium depletion boundary and corresponding age for the BPMG is presented in Section~\ref{ldb}.

Three of our candidates have strong \ha\ in emission and  Li absorption yet do not have RVs consistent with BPMG membership.  We have designated these as possible members (with `Y?') they may be SB1s where an unseen companion is causing the RV deviation. Alternatively, these are young field interlopers into our sample.

\section{Summary of Confirmed BPMG Members 
}\label{members}

The methodology of the ACRONYM of the BPMG is summarized in Figure~\ref{flowchart}, which traces the prescription of our candidate selection and confirmation.  We combine proper motion limits, suggestion of pre-main-sequence luminosity through photometric distance constraints, spectroscopic youth indicators such as low gravity, Li and \ha, and RVs for kinematic matching.

Of our 104 BPMG candidates with follow-up IR and/or optical spectroscopy, we confirmed 66 member systems, recovering 25 previously reported members and identifying 41 new ones. Nineteen of the 66 members have spectroscopic or visual companions.  All of our candidates with their final membership status are listed in Table~3. 
Fourteen of the 66 members have `Y?' status.  Ten of the candidates are SBs with system RVs consistent with the BPMG, strong \ha\ emission, with two of these with strong Li absorption confirming youth.  Eight of the SBs have inconclusive Li observations, but to avoid biasing the sample against SBs, we designate these with no Li information as `Y?'. These systems should be followed up for full orbital solutions and/or parallaxes to better understand their membership status.  

Our confirmation yield for BPMG systems is 66 of 104 candidates (63\%), a little lower than the 67\% confirmation rate we achieved in our Tuc-Hor study (\citealt{krau14}). The slightly lower yield results from the extension of our search to positions further north than known BPMG members.  Though this extension decreased our spectroscopic yield, it allowed us to determine the northern boundary of the BPMG (Figure~\ref{bpmgmap}). 

\begin{figure}[tbp]
\includegraphics[width=6.5in,angle=0]{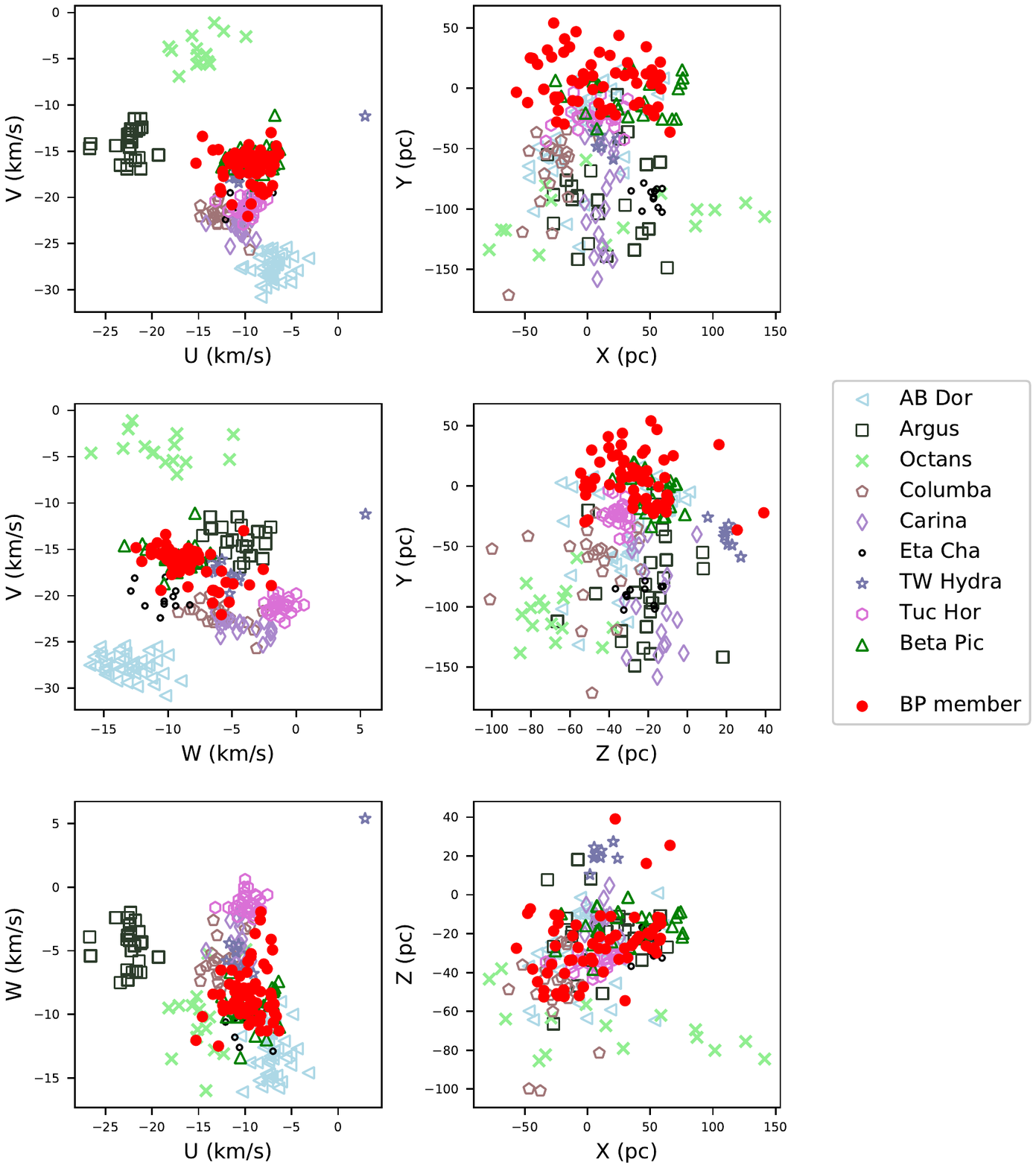}
\vspace{-1in}
	\caption{UVW and XYZ of known members of nine YMGs, including the new BPMG members reported in this work (red circles). The few outliers in U,W plot in the bottom left corner have `Y?' based on known binarity, no lithium detection, or large RV uncertainty. The other YMG members are from \cite{torr08}.
	\label{uvwxyz}}
\end{figure}

The space velocities (UVW) and positions (XYZ) of confirmed members were calculated using the positions, RVs, proper motions, and kinematic distances. They are compared to the YMGs in Figure~\ref{uvwxyz} and are listed in Table 3. 
The few outliers in these figures are the known or suspected SBs. Of the kinematic matches to the BPMG, only two are clearly not members, implying a 3\% false-membership rate based on kinematics alone. This number is slightly less than that calculated by \cite{shko12}, who measured the UVWs of all the nearby old M dwarfs and found that 6\% have kinematic matches to BPMG, but are clearly old. Our process for photometrically selecting pre-main sequence stars improves our ability to use kinematics to identify new candidate members.

To present the new mass function for the BPMG, we converted the SpT$_{SED}$ for each target at an assumed age of 23 Myr to mass using the SpT--T$_{eff}$ relation from \cite{herc14} for SpTs of F5 and later and from \cite{peca13} for stars earlier than F5. We then used the mass--T$_{eff}$ relation from \cite{bara15} for stars with masses $\leq$1.4 \Msun, and the latest PARSEC models of \cite{chen15} for masses $>$1.4 \Msun. 

\begin{figure}[tbp]
\center
\includegraphics[width=4.5in,angle=0]{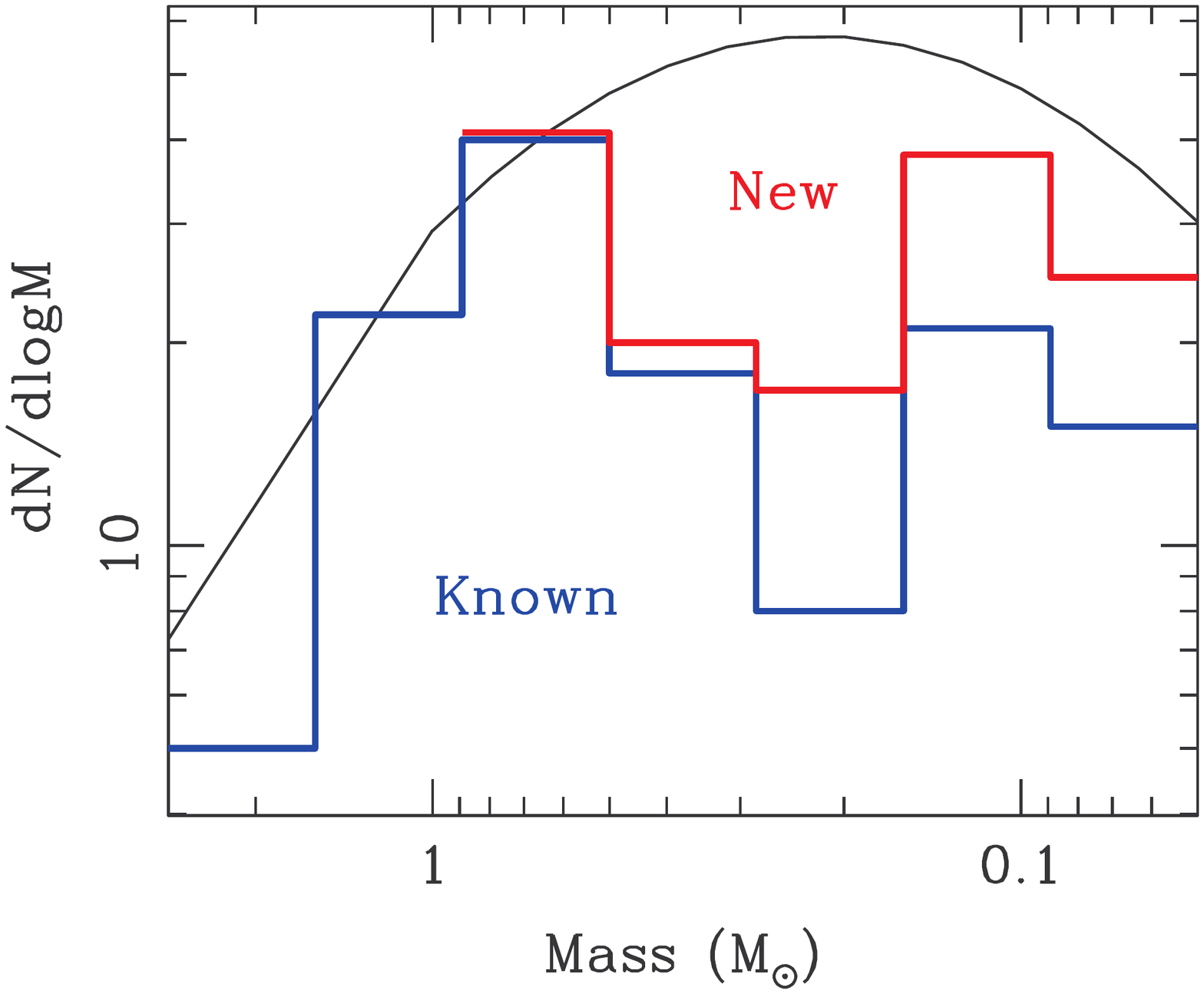}
	\caption{Mass histogram of BPMG member systems. The previously published members are shown in blue with the addition of those found in this study shown in red. The new additions add 1, 2, 7, 17, and 10 objects, respectively, in the lowest five bins. These increase the number of known M stars in the BPMG by 46\% from 80 to 117. The black line shows the Salpeter+Chabrier IMF normalized to the most populated mass bin, indicating that there are a significant number of M stars remaining to be discovered. There are also 14 literature substellar objects not shown on this mass scale to which we add four more. We did not attempt to estimate the masses of those targets as they have SpTs later than M7 and the models for these have not yet been empirically calibrated.
	\label{massfunction}}
\end{figure}

In Figure~\ref{massfunction} we compare the mass histogram for the BPMG before and after our confirmation of the new low-mass members. The new additions add 1, 2, 7, 17, and 10 objects respectively, in the lowest five mass bins.  There are also 14 literature substellar objects, to which we add four more, for which we did not attempt to estimate the masses as they have SpTs later than M7 and the models for these have not yet been empirically calibrated.

\subsection{Lithium Depletion and the Age of BPMG}\label{ldb}

Low-mass stars are gradually depleted of lithium as convection carries
it deep into the stellar interior, where fusion converts it to helium.
Lithium burning proceeds most rapidly for early-M stars (e.g., Figure~\ref{spt_halpha_lithium}, right), while stars of higher and lower mass burn lithium at
correspondingly slower rates. This transition from lithium-rich to
lithium-depleted is particularly rapid at the lower end of the
depleted range, leading to a narrow ``lithium depletion boundary''
(LDB) that serves as a sensitive probe of age. The BPMG is one of the
youngest nearby moving groups for which lithium depletion of early M
dwarfs has be observed (e.g., \citealt{ment08}), leading to a lithium
depletion age of $\tau = 24 \pm 4$ Myr (Binks \& Jeffries 2016). The
paucity of known M3--M5 members of BPMG has led to uncertainty in the
location and width of the LDB, and hence the moving group's age, but
our newly identified BPMG members now densely sample this mass range.

\begin{figure}[tbp]
\includegraphics[width=5.5in,angle=180]{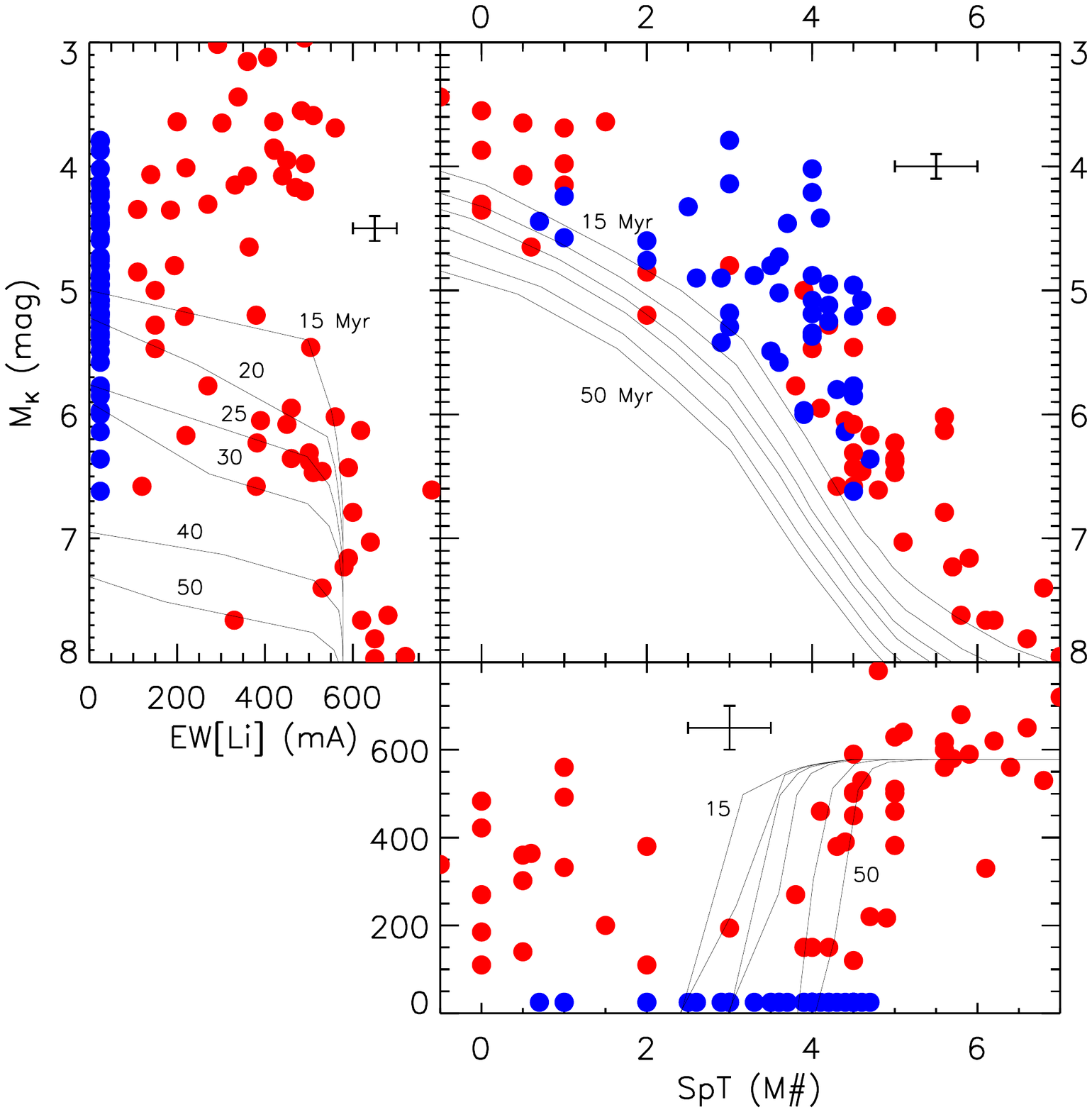}
	\caption{SpT$_{SED}$, absolute $M_K$ magnitude, and
lithium equivalent width of all of the K7--M6 members of the BPMG, as
projected onto the three corresponding two-dimensional planes. The
BPMG members without lithium (blue) are clearly clustered among the M1--M4
stars, with most $\le$M0 and $\ge$M4 stars bearing at least some
lithium (red). Representative error bars are shown on each plot.  The isochronal tracks of Baraffe et al. (2015) are shown in each
plane, spanning ages of 15--50 Myr. We converted the isochrones into
observational space by adopting their estimate of $M_K$, applying the
SpT-$T_{eff}$ temperature scale of Herczeg \& Hillenbrand (2014), and
applying the $A_{Li}$-$EW[Li]$ curves of growth of Palla et al.
(2007).
 \label{BPMGLi}}
\end{figure}

In Figure \ref{BPMGLi}, we plot the SpT$_{SED}$, absolute $M_K$ magnitude, and
lithium equivalent width of all of the K7--M6 members of the BPMG, as
projected onto the three corresponding two-dimensional planes. The
BPMG members without lithium are clearly clustered among the M1--M4
stars, with most $\le$M0 and $\ge$M4 stars bearing at least some
lithium. To provide context for the pattern of lithium depletion, we
also show the isochronal tracks of Baraffe et al. (2015) in each
plane, spanning ages of 15--50 Myr. We converted the isochrones into
observational space by adopting their estimate of $M_K$, applying the
SpT-$T_{eff}$ temperature scale of \cite{herc14}, and
applying the $A_{Li}$-$EW[Li]$ curves of growth of \cite{pall07}.

Our results clearly demonstrate the tension between theory and
observations for age determinations of young stars. In the HR diagram,
nearly all sample members fall above the 15 Myr isochrone, implying a
very young age of $\tau \sim 10$ Myr. However, the cluster sequence in
$M_K$ vs $EW[Li]$ implies an older age of $\tau \sim$25 Myr, and the
corresponding sequence in SpT$_{SED}$ vs $EW[Li]$ implies an even older age of
$\tau \ga$50 Myr. HR diagram ages are now broadly accepted to
underestimate the true age of stars (e.g., \citealt{nayl09,peca12}), most likely due to theoretical overestimation of
$T_{eff}$ for a given mass (e.g., \citealt{krau15}). However, theory
and observations seem to broadly agree on the relation between
luminosity and mass for M stars at 10--50 Myr (\citealt{krau15,mont15,rizz16,niel16}). The
model-derived ages in our three observational planes are consistent
with these patterns, so we hereafter refine our analysis to the
($EW[Li]$,$M_{K}$) plane of Figure~\ref{BPMGLi}.

The observed sequence broadly agrees with the results of \cite{bink16}, who estimated the LDB age to be $\tau = 21 \pm 4$ Myr
from the models of \cite{bara15} or $\tau = 24 \pm 4$ Myr from
the magnetized models of \cite{feid16}. Most of our observed
lithium-bearing stars do indeed fall between the 20 and 25 Myr
isochrones. However, \cite{bink16} identified the LDB primarily via a discrete
gap between the lithium-bearing and lithium-depleted members, while
our more dense sequence also demonstrates for the first time that
there is a finite width in SpT to the LDB. There is no clear gap in
luminosity where there are no members with lithium, while
lithium-depleted members extend as faint as $M_K = 6.6$. If we isolate
the analysis to $\ge$M2 stars and identify the boundary to be the
$M_K$ where there are an equal number of lithium-depleted and lithium-rich members, then we find a
boundary location of $M_K = 5.6$ mag, with 8 stars in each group. We find that 68\% of the
interlopers fall in the range of $5.2 < M_K < 6.0$ mag, so the
corresponding uncertainty is $\pm$0.4 mag. The corresponding age from
the  \cite{bara15} models is $\tau = 22 \pm 6$ Myr.

Unresolved binarity could explain some extension of the
lithium-bearing population upward, but it is difficult to explain the
faintness of some lithium-depleted members; even the faintest
lithium-depleted member (2MASS J03255277-3601161) shows excellent
agreement with BPMG membership in all other aspects. The observations
could indicate a genuine age spread, as has been suggested for AB Dor
(\citealt{lope06}) and for Taurus-Auriga/32 Ori (\citealt{krau17}). However, it also might result from variations in
lithium burning rates due to other stellar parameters. Lithium
abundances have long been known to correlate with rotation for more
massive stars (e.g., \citealt{sode93}), including in the BPMG
(\citealt{mess16}), and substantial variations in lithium abundance
have also been seen in populations both younger (Upper Sco; \citealt{rizz15}) and older (Pleiades; \citealt{some17}). Convective
mixing could be correlated with rotation at varying levels of
subtlety, from direct impact on the convective mixing length (\citealt{bouv08}), to magnetic activity suppressing convection (\citealt{chab07,feid16}), or even more subtly due to delayed rise in the
core temperature (\citealt{bara10,some14}). There is no straightforward way to disentangle these effects until
purely geometric ages from astrometric tracebacks become feasible. However, our measured width for the LDB ($M_K =
5.6 \pm 0.4$ mag) is three times broader than the corresponding boundary
we measured in Tuc-Hor ($M_K = 7.12 \pm 0.16$) via identical
techniques. Either the astrophysics broadening the sequence change
between 20 Myr and 40 Myr, or the BPMG may indeed have a substantially
larger age spread than Tuc-Hor.

Finally, we find two outliers for which their
membership should be further inspected:
\begin{itemize}
    \item 2MASS J03363144-2619578 (SpT$_{SED}$=M6.1; $M_K = 7.66$) appears to be
partially depleted in lithium ($EW[Li] = 330$ m\AA), even though it
should be undepleted according to both SpT and $M_K$. It also has been
suggested as a candidate member of the Tuc-Hor moving group (\citealt{rodr13,gagn15b}), and its observed RV (= 16.7 $\pm$ 2.7 km/s) is only marginally more consistent with expectations
for the BPMG (= 18.2 km/s) than for Tuc-Hor (= 13.8 km/s). The lithium abundance suggests that the older age of Tuc-Hor is
more appropriate, but its membership will ultimately be confirmed by
parallax. It should fall at $d = 25$ pc if in the BPMG or at $d = 50$ pc
if in Tuc-Hor.

\item 2MASS J02365171-5203036 (=GSC 08056-00482) was proposed by \cite{zuck04} to be an M3 member of Tuc-Hor, but \cite{elli14}
subsequently classified it as an M2 member of the BPMG, and \cite{malo14} used their BANYAN formalism to estimate a most likely
membership in Columba, albeit with a significant probability for BPMG
and a small probability for Tuc-Hor. This star sits low on the cluster
sequence for BPMG, but its high lithium abundance for its SpT ($EW[Li] = 380$~m\AA; \citealt{torr06}) would be inconsistent with Tuc-Hor or
Columba membership for M2--M3 stars (e.g., Kraus et al. 2014), and is
high even for the BPMG. Membership in the BPMG seems most likely, but a
parallax would clarify the question of membership for this star.

\end{itemize}

\section{Conclusion}

In this paper we present the results from our efficient photometric+kinematic selection process, which identified 104 low-mass BPMG candidates. Follow-up infrared observations of the latest spectral types (SpT $>$ M5) were collected with IRTF/SpeX to search for low gravity objects, an indication of youth.  We observed the young ones of these, plus all the earlier SpT candidates with the high-resolution optical spectrographs at the Magellan and Keck telescopes.  These data provided  the RVs needed to confirm or reject stars co-moving with the BPMG and youth indicators such as Li absorption and \ha\ emission.

Prior to our work, there were 94 known member systems with primary masses less than 0.7 $\Msun$. We compiled the current census in the Appendix with which we were able to test the efficiency and false-membership assignments using our selection and confirmation criteria.  The addition of 41 new systems from this work represents a 44\% increase in low-mass BPMG membership and a new sample around which to look for directly-imaged planets. The four new BPMG members with SpTs $\geq$ M7, and likely less than 0.07 $\Msun$, represents a 29\% increase to the 14 already known. They are: 
2MASS J00413538-5621127,
2MASS J03550477-1032415,
2MASS J22334687-2950101, and
2MASS J23010610+4002360.
If the number of BPMG's AFGK stars is complete, then the fraction of M stars in the BPMG is now 69\% nearing the expected value of 75\%.
However, the new IMF for the BPMG clearly shows a deficit of 0.2-0.3\Msun\ stars by a factor of $\sim$2. We except that the AFGK census  of the BPMG is \textit{also} incomplete, probably due to biases of searches towards the nearest of stars.  Future surveys for AFGK and M spectral type members should extend out to at least 100 pc for the BPMG.


\section{Appendix: Members of BPMG From the Literature}\label{appendix}

To complement our search for new members, we compiled the most comprehensive list of the 146 objects that have been reported in the literature as likely members of the BPMG in Table 4. This list includes every object that has been asserted to be a member, as well as any probabilistic candidates  that have  $>50\%$ membership probability (e.g., \citealt{malo13}). Past stars considered members that were later refuted by more data are not included (\citealt{bink16,elli16,liu16}.)
We also exclude any purported members from traceback analyses \citep{orte02,song03,maka07} as recent revisions to the age of the BPMG cast doubt on tracebacks conducted with the previous (younger) age \citep{mama14}. In compiling our census, we included every object that has a distinct entry in the 2MASS Point Source Catalog (\citealt{cutr03}), but did not include secondary components of close multiple systems. The effective angular separation cutoff is $\approx$4--5\arcsec. We list the previously reported members in Table 4 along with all available data used for membership assessments: spectral types, proper motions, RVs, H$\alpha$ EWs, Li EWs, and surface gravity assessments. We also compiled parallaxes for 77 of the objects in the literature census. Parallaxes are not currently available for any sources in our own survey.  For easier use of the complete census, we appended the new BPMG members from this work to the bottom of Table 4. 

We subjected the 113 published members with K and M spectral types to the same tests described in Sections 3--5 as a test of the methodology.  The results of those tests are summarized below:

\begin{itemize}

\item  88 have membership designations of `Y' (77.9\%).

\item 17 have `Y?' (15.0\%) because the star's RV=`?' or `N', but it has \ha=`Y', Li=`Y' and/or has been shown to have low surface gravity.

\item 6 have membership=`N' (5.3\%): \\
		$\diamond$ 4 of these have RV=`Y' but membership=`N' because \ha=`N' and Li=`?'. These cases may be the few relatively \textit{inactive} young stars that may exist, or are not in fact members with kinematics coincident with the BPMG.  \\
		$\diamond$ 2 have RV=`?', Li=`?', \ha=`?' \textit{and} field gravities from \cite{gagn15b} and are probably not BPMG members.
		
\item 1 has membership=`?' (0.9\%) because it has no RV, \ha, or Li information, and is reported to have low surface gravity.

\end{itemize}

Assuming all the literature-reported members in Table 4 are indeed members, below are fractions that do not pass each test, representing a possible false-negative rate for each age diagnostic:\footnote{Twenty-six of 74 (35.1\%) stars have no measurable Li absorption, but this is not an indication of old age for BPMG members (Figure~\ref{spt_halpha_lithium}). }
\begin{itemize}
    
\item 7 of 100 (7.0\%) with published RV do not have velocities consistent with membership. 
\item 4 of 96 (4.2\%) with measured \ha\ have lower than expected emission for young stars.
\item 2 of 12 (17\%) with reported gravity classification have field level gravity and thus likely are not young.  

\end{itemize}

In conclusion, we propose that these false-negative rates be considered  when confirming new members, but caution should be applied to not use these in a circular fashion to increase our tolerance of each age diagnostic.

\acknowledgements

E.S.~thanks J.~Teske for useful discussions and appreciates support from the HST grant HST-AR-13911.002-A and NASA/Habitable Worlds grant NNX16AB62G. K.A. acknowledges support by a NASA Keck PI Data Award (\#1496879), administered by the NASA Exoplanet Science Institute. We also thank Dr.~Eric Mamajek for a speedy and helpful referee's report. This research has made use of the VizieR catalogue access tool, CDS, Strasbourg, France \citep{ochs00} and the Mikulski Archive for Space Telescopes (MAST). STScI is operated by the Association of Universities for Research in Astronomy, Inc., under NASA contract NAS5-26555. Support for MAST for non-HST data is provided by the NASA Office of Space Science via grant NNX13AC07G and by other grants and contracts.


\newpage


\begin{deluxetable}{lllll l lll ll l}\label{table_candidates}
	\rotate
 \tabletypesize{\tiny}
 \tablewidth{0pt}
 \tablecaption{Selection Criteria and Observations of Candidate BPMG Members}
 \tablehead{
 \colhead{2MASS J} & \colhead{$R_{USNOB}$} & \colhead{$K_s$} & \colhead{$\mu$} & \colhead{SpT$_{SED}$} & \colhead{$m_{bol}$} & \colhead{$DM_{phot}$} & \colhead{$\Delta_{PM}$} & \colhead{$DM_{kin}$} & \colhead{IR Spec.\tablenotemark{a}} & \colhead{Opt. Spec.} & \colhead{UT date} 
 \\
 \colhead{} & \colhead{(mag)} & \colhead{(mag)} & \colhead{(mas/yr)} & \colhead{} &  \colhead{(mag)} & \colhead{(mag)} & \colhead{(mas/yr)} & \colhead{(mag)} & \colhead{} &  \colhead{} & \colhead{(yyyymmdd)}
  }
 \startdata
00140580-3245599 & 18.86 & 12.03 & (73,-46.3) $\pm$ 9.1 & M7.7 $\pm$ 0.2 & 15.02 $\pm$ 0.02 & 2.02 $\pm$ 0.07 & 10.2 & 3.4 & SXD & -- & -- \\
00164976+4515417 & 14.68 & 9.85 & (62.5,-22.1) $\pm$ 3.5 & M4.5 $\pm$ 0.3 & 12.61 $\pm$ 0.03 & 2.14 $\pm$ 0.25 & 9.6 & 4.0 & -- & Keck/HIRES & 20120710 \\
00172353-6645124 & 11.51 & 7.70 & (104.3,-13.5) $\pm$ 1.0 & M2.6 $\pm$ 0.4 & 10.31 $\pm$ 0.02 & 1.48 $\pm$ 0.24 & 2.4 & 2.8 & -- & Magellan/MIKE & 20101229 \\
00193931+1951050 & 14.39 & 10.08 & (72.2,-49.7) $\pm$ 3.5 & M4.5 $\pm$ 0.1 & 12.82 $\pm$ 0.02 & 2.36 $\pm$ 0.10 & 2.9 & 3.5 & SXD & Magellan/MIKE & 20101231 \\
00194303+1951117 & 14.19 & 9.86 & (70.6,-50.9) $\pm$ 3.5 & M4.7 $\pm$ 0.1 & 12.64 $\pm$ 0.02 & 1.96 $\pm$ 0.10 & 4.9 & 3.5 & SXD & Magellan/MIKE & 20101231 \\
00233468+2014282 & 9.97 & 7.34 & (61.4,-38.3) $\pm$ 1.9 & K7.6 $\pm$ 0.2 & 9.65 $\pm$ 0.04 & 2.34 $\pm$ 0.11 & 1.5 & 3.9 & -- & Keck/HIRES & 20120710 \\
00274534-0806046 & 16.46 & 10.61 & (97.3,-66.9) $\pm$ 3.3 & M6.6 $\pm$ 0.1 & 13.55 $\pm$ 0.02 & 1.10 $\pm$ 0.06 & 6.6 & 2.8 & SXD & Keck/HIRES & 20120710 \\
00281434-3227556 & 13.97 & 9.28 & (108.0,-42.6) $\pm$ 2.9 & M4.3 $\pm$ 0.4 & 12.01 $\pm$ 0.04 & 1.76 $\pm$ 0.34 & 5.7 & 2.7 & -- & Keck/HIRES & 20120710 \\
\enddata
\tablenotetext{a}{IRTF's SXD mode: 0.8-2.4 \micron.  See Table 2 for log of observations.}
\tablenotetext{b}{NIR SpT indicates M9 \citep{alle13}.}
\tablenotetext{c}{This object was observed with SXD by \cite{alle13}.}
\tablenotetext{d}{Optical spectrum appears as early-type SB2, not an M star.}

 \end{deluxetable}

  \begin{deluxetable}{lcccllcl}\label{table_irdata}
 \tablewidth{0pt}
 \tabletypesize{\tiny}
 \tablecaption{NIR Spectroscopy Log}
 \tablehead{
 \colhead{Object} & \colhead{UT date} & \colhead{Slit Width} & \colhead{sec $z$} & \colhead{$N_{exp} \times t \mathrm{(s)}$} & \colhead{S/N} & \colhead{SpT$_{NIR}$\tablenotemark{a}} & \colhead{Gravity\tablenotemark{a}} \\
 \colhead{2MASS J} & \colhead{(YYYY-MM-DD)} & \colhead{($\arcsec$)} & \colhead{} & \colhead{} & \colhead{(J,H,K)} & \colhead{} & \colhead{Type} 
  }
 \startdata 
00140580-3245599& 2010-11-14& 0.8&2.23&8~$\times$~90.0&132, 118, 134&     M7& {\sc fld-g}\\
  00193931+1951050& 2011-07-07& 0.3&1.21&10~$\times$~30.0&44, 49, 42&     M4&     \nodata\\
  00194303+1951117& 2011-07-07& 0.3&1.18&8~$\times$~45.0&68, 74, 64&     M4&     \nodata\\
  00274534-0806046& 2017-01-05& 0.8&1.19&12~$\times$~29.7&47, 60, 68&     M6& {\sc fld-g}\\
  00570256-1425174& 2010-11-11& 0.8&1.22&16~$\times$~60.0&125, 117, 93&     M6& {\sc fld-g}\\
  01351393-0712517& 2010-11-15& 0.3&1.12&4~$\times$~30.0&79, 57, 38&     M4&     \nodata\\
  02241739+2031513& 2010-10-05& 0.8&1.03&6~$\times$~60.0&106, 111, 104&     M6& {\sc int-g}\\
  02485260-3404246& 2010-11-15& 0.3&1.69&6~$\times$~60.0&73, 49, 30&     M5& {\sc int-g}\\
\enddata
  \tablenotetext{a}{Stars with SpTs earlier than M4 are outside the valid range of the AL13 NIR spectral typing method. Uncertainty in the IR spectral types is $\pm$ 1 subclass as discussed in AL13. Stars with SpTs earlier than M5 cannot be classified with the AL13 NIR gravity classification. }
  \end{deluxetable}

 
   \begin{deluxetable}{lcllllllllllll}\label{table_membership}
	   \rotate
  \tabletypesize{\tiny}
  \tablewidth{0pt}
  \tablecaption{Results of Optical Spectroscopy and Membership Assessments}
  \tablehead{
  \colhead{Name} &  \colhead{SpT} & \colhead{SpT} &\colhead{RV\tablenotemark{a}} & \colhead{$\Delta$RV\tablenotemark{b}} & \colhead{Binary?} &  \colhead{$EW[Li]$\tablenotemark{d}} & \colhead{$EW[H\alpha]$\tablenotemark{d}}  & \colhead{UVW\tablenotemark{e}} & \colhead{XYZ\tablenotemark{e}} & \colhead{RV, \ha, Li \tablenotemark{f}}  & \colhead{BPMG} & \colhead{Prev.} 
  \\
 \colhead{} &   \colhead{$Spec$} & \colhead{Source (ref)} & \colhead{(km/s)} & \colhead{(km/s)} & \colhead{(Ref)\tablenotemark{c}} & \colhead{(\AA)} & \colhead{(\AA)} & \colhead{(km/s)}  & 
 \colhead{(pc)} & \colhead{Assessment} & \colhead{Memb.?} & \colhead{Ref.\tablenotemark{g}} 
  } 
  \startdata
00140580-3245599	&	M7	&	IR (A)	&	--	&	--	&	--	&	--	&	--	&	--	&	--	&	???	&	N	&	\\				
00164976+4515417	&	M4.1	&	Opt (B)	&	-12.05	$\pm$	0.64	&	-5.5	&	VB(1)	&	$<$0.05	&	-7.29	&	(-9.4,-20.7,-5.2)	&	(-27,53.9,-18.6)	&	Y$^?$Y?	&	Y?	&	\\		
00172353-6645124	&	\textit{M2.5}	&	Opt (C)	&	9.50	$\pm$	0.90	&	-1.58	&	--	&	$<$0.05	&	-4	&	(-11.1,-15.2,-8.2)	&	(14.4,-18.3,-27.8)	&	YY?	&	Y	&	1\\		
00193931+1951050	&	M4	&	IR (A)	&	-6.98	$\pm$	0.37	&	-5.25	&	VB	&	0.12	&	-6.77	&	(-8,-19.7,-6)	&	(-14.3,34.1,-33.8)	&	YYY	&	Y?	&	\\		
00194303+1951117	&	M4	&	IR (A)	&	-6.66	$\pm$	0.26	&	-4.93	&	VB	&	$<$0.05	&	-5.13	&	(-7.6,-19.4,-6.4)	&	(-14.3,34.1,-33.8)	&	YY?	&	Y	&	\\		
00233468+2014282	&	K7.5	&	Opt (B)	&	-2.17	$\pm$	0.59	&	-0.64	&	VB	&	0.36	&	-0.18	&	(-10.1,-16.2,-8.5)	&	(-18.3,40.8,-40.5)	&	YYY	&	Y	&	2\\		
00274534-0806046	&	M5.6	&	Opt (B)	&	-3.61	$\pm$	1.02	&	-7.97	&	--	&	0.65	&	-5.53	&	(-8.3,-18.9,-1.9)	&	(-3.3,11.9,-34.1)	&	NYY	&	Y?	&	3\\		
00281434-3227556	&	M4.8	&	Opt (B)	&	6.79	$\pm$	2.66	&	-1.62	&	VB(2)	&	0.38	&	-10.63	&	(-11,-15.3,-7.7)	&	(4.3,-1.2,-34.4)	&	YYY	&	Y	&	\\
\enddata
\end{deluxetable}

   \begin{deluxetable}{llrlrrrrrllllllll}\label{knownmem}
	   \rotate
  \tabletypesize{\tiny}
  \tablewidth{0pt}
  \tablecaption{Known Members from the Literature + New Members from This Work}
  \tablehead{
  \colhead{2MASS J} & \colhead{Other} & \colhead{Ref} & \colhead{SpT} & \colhead{Ref} & \colhead{EW[H$\alpha$]} & \colhead{Ref} & \colhead{EW[Li]} & \colhead{Ref} & \colhead{RV} & \colhead{Ref} & \colhead{Grav.} & \colhead{Ref} & \colhead{$\pi$} & \colhead{Ref}  & \colhead{RV, \ha, Li}  & \colhead{BPMG}
  \\
 \colhead{} & \colhead{Name} & \colhead{} & \colhead{} & \colhead{} & \colhead{(\AA)} & \colhead{} & \colhead{(m\AA)} & \colhead{} & \colhead{(km/s)} & \colhead{} & \colhead{Type} & \colhead{} & \colhead{(mas)} & \colhead{} & \colhead{Assessment} & \colhead{Memb.?}\\
 \colhead{} & \colhead{} & \colhead{} & \colhead{} & \colhead{} & \colhead{} & \colhead{} & \colhead{} & \colhead{} & \colhead{} & \colhead{} & \colhead{} & \colhead{} & \colhead{} & \colhead{} & \colhead{for Ks \& Ms\tablenotemark{a}} & \colhead{for Ks \& Ms\tablenotemark{a}}
  } 
  \startdata
  \textbf{Known Members}\\
\hline
00065008-2306271 & HD 203 & 1 & F2 & 1 & .. & .. & 87 & 31 & 9.7 $\pm$ 2.0 & 31 & .. & .. & 25.2 $\pm$ 0.6 & 65 & & \\
00172353-6645124 & HIP 560 & 2 & M2.5      & 5 & -6.0 & 34 & .. & .. & 10.7 $\pm$ 0.2 & 5 & .. & .. & 25.6 $\pm$ 1.7 & 17 &YY-- &Y \\
00233468+2014282 & TYC 1186-706-1 & 3 & K7.5 & 3 & -0.2 & 44 & 339 & 21 & -1.7 $\pm$ 0.2 & 5 & .. & .. & .. & .. &YYY &Y \\
00274534-0806046 &  & 4 & M5.5 & 4 & .. & .. & .. & .. & .. & .. & INT-G & 4 & .. & .. &-- -- -- &? \\
00275023-3233060 & GJ 2006 A & 5 & M3.5     & 5 & -7.8 & 34 & .. & .. & 8.8 $\pm$ 0.2 & 5 & .. & .. & 30.1 $\pm$ 2.5 & 17 &YY-- &Y \\
00275035-3233238 & GJ 2006 B & 5 & M3.5     & 5 & -5.3 & 34 & .. & .. & 8.5 $\pm$ 0.2 & 5 & .. & .. & 31.8 $\pm$ 2.5 & 17 &YY-- &Y \\
00325584-4405058 &  & 6 & L0 & 26 & .. & .. & .. & .. & 13.0 $\pm$ 1.9 & 55 & INT-G & 26 & 38.4 $\pm$ 4.8 & 66 & & \\
\enddata
  \tablenotetext{a}{The last two columns list the assessment of each age diagnostic and membership status following the prescription of this paper as summarized by Figure~\ref{flowchart}.  We subjected the 113 published members with K and M spectral types to the same tests described in Sections 3--5.  The results of those tests are summarized in the text.}
  \tablecomments{Refs:
  1: \cite{zuck01}, 2: \cite{malo13}, 3: \cite{lepi09}, 4: \cite{gagn15b}, 5: \cite{malo14}, 6: \cite{gagn15a}, 7: \cite{kiss11}, 8: \cite{bink16}, 9: \cite{elli16}, 10: \cite{schl10}, 11: \cite{song03}, 12: \cite{moor06}, 13: \cite{alon15}, 14: \cite{elli14}, 15: \cite{rodr14}, 16: \cite{torr08}, 17: \cite{ried14}, 18: \cite{schl12a}, 19:\cite{teix09}, 20: \cite{schl12b}, 21: \cite{malo14b}, 22: \cite{moor13}, 23: \cite{liu16}, 24: \cite{barr99}, 25: \cite{liu13}, 26: \cite{alle13}, 27: \cite{bink14}, 28: \cite{gray03}, 29: \cite{lope10}, 30: \cite{zuck04}, 31: \cite{torr06}, 32: \cite{kord13}, 33: \cite{alca00}, 34: \cite{riaz06}, 35: \cite{houk88}, 36: \cite{shko09b}, 37: \cite{teix09}, 38: \cite{step86}, 39: \cite{hawl97}, 40: Hipparcos Catalog, 41: \cite{reid08}, 42: \cite{cruz09}, 43: \cite{shko12}, 44: \cite{lepi13}, 45: \cite{reid07b}, 46: \cite{alca96}, 47: \cite{whit07}, 48: \cite{song12}, 49: \cite{herb88}, 50: \cite{mart10}, 51: \cite{gaid14}, 52: \cite{reid02a}, 53: \cite{dasi09}, 54: \cite{krau14}, 55: \cite{fahe16}, 56: \cite{mont01}, 57: \cite{bail12}, 58: \cite{maci15}, 59: \cite{gizi02b}, 60: \cite{gont06}, 61: \cite{stra00}, 62: \cite{wils53}, 63: \cite{zuck00}, 64: \cite{alle16a}, 65: \cite{gaia16}, 66: \cite{fahe13}, 67: \cite{vanl07}, 68: \cite{schn17}, 69: this work (see Tables 1-3 for more details on these members.  }

  \end{deluxetable}

\bibliography{refs_master_2017June}{}
\bibliographystyle{apj}

\end{document}